\documentstyle[12pt,aasms4]{article}
\newcommand{\degrees}{\mbox{$^{\circ}$}}    
\newcommand{\chisq}{\mbox{$\chi^2$}}       
\newcommand{\chinu}{\mbox{$\chi_{\nu}^2$}}  
\newcommand{\etal}{{\rm et al.}}           
\newcommand{\NH}{\mbox{${\rm N_H}$}}       
\newcommand{\NHunits}{\mbox{$ 10^{20}~{\rm cm}^{-2}$}}

\newcommand{\pks}{\mbox{\rm PKS 1510-089}}       

\begin{document}
 
\title {X-RAY SPECTRUM OF THE HIGH POLARIZATION QUASAR PKS~1510-089}
 
\author {K. P. Singh \altaffilmark{1, 4},
         C. R. Shrader \altaffilmark{2, 3, 5}, 
         and I. M. George \altaffilmark{2, 3, 6}}

\altaffiltext{1} {X-ray Astronomy Group, Tata Institute of Fundamental
Research,
 Mumbai 400 005, India}
\altaffiltext{2}{Laboratory for High Energy Astrophysics, NASA/GSFC,
 Greenbelt, MD 20771}
\altaffiltext{3}{Also Universities Space Research Association, 7501
Forbes Blvd, 
Seabrook MD  20706}
\altaffiltext{4}{singh@tifrvax.tifr.res.in}
\altaffiltext{5}{shrader@grossc.gsfc.nasa.gov}
\altaffiltext{6}{ian.george@gsfc.nasa.gov}

\slugcomment{To appear in {\em The Astrophysical Journal}, 1997 December 1  }

\begin{abstract}

We present results on the X-ray spectra of the radio-loud, 
high-polarization quasar, \pks, based on new data obtained using 
$ASCA$, and from archival $ROSAT$ data.
We find the X-ray spectrum obtained by $ASCA$ to be unusually hard,
with the photon index, $\Gamma$=1.30$\pm$0.06, while the
(non-simultaneous)
$ROSAT$ data indicate a steeper spectrum with $\Gamma$=1.9$\pm$0.3.  
However, we find the X-ray flux at 1 keV to be within $\sim$10\% during
both 
observations. Thus we suggest the most likely explanation is that 
there is a break (with $\Delta \Gamma \geq 0.6$) 
in the underlying continuum at $\sim$~0.7~keV.

Although the sample of high-polarization quasars for which high quality 
X-ray spectra are available is small, flat X-ray spectra seem to be the 
characteristic of these objects, and they also appear to be harder than 
that of the other radio-loud but low-polarization quasars. 
The multiwavelength spectrum of \pks~is similar to many 
other $\gamma$-ray blazars, suggesting the emission is dominated 
by that from a relativistic jet.  A big blue-bump is also seen 
in its multiwavelength spectrum, suggesting the presence of a
strong thermal component as well.

\end{abstract}
 

\section{INTRODUCTION}
High polarization quasars (HPQs), along with BL Lac objects, form a
subset of 
active galactic nuclei (AGN) known as blazars. HPQs are among the most
energetic AGN.  
They are characterized by strong and variable radio, infrared and 
optical continuum emission with a high degree of polarization. Blazars 
are also strong and variable soft X-ray emitters (Brunner et al. 1994), 
and recently have also been found to emit $\geq$100 MeV $\gamma$-rays 
(Thompson \etal 1993).  The X-ray spectral measurements of blazars have
shown that the HPQs tend to have flatter spectral indices than the 
BL Lac objects (Sambruna \etal 1994).  The uncertainties in these
measurements of spectral indices are, however, generally large due 
to the small bandwidth, poor sensitivity, and low-energy absorption 
in the interstellar medium.
There is also a general dearth of broad-band spectral 
measurements of HPQs extending to hard X-rays, the few exceptions being
3C~279 and 3C~345 (Makino 1989), \pks~(Singh, Rao, \& Vahia 1990), 
PKS~0537-441 (Sambruna et al. 1994), 3C~390.3 (Eraclous, Halpern, \& 
Livio 1996), and a serendipitous detection of PKS~1502+106 
(George \etal 1994).  
We chose to carry out X-ray spectral measurements of PKS~1510-089 
with the $ASCA$ observatory as previous measurements with the $EXOSAT$
Observatory, reported by Singh \etal (1990) and Sambruna \etal (1994), 
indicated an unusually flat X-ray spectrum.

 Radio source PKS~1510-089 was first identified optically as a quasar 
with an ultraviolet excess, a visual magnitude of 16.5 (Bolton \& Ekers 
1966), and a redshift of $z=0.361$ measured from its emission line
spectrum 
(Burbidge \& Kinman 1966).
Optical polarimetric observations by Appenzeller \& Hiltner (1967)
detected
strong (10.9\% $\pm$ 3.8\%) polarization at a position
angle of 177$^{\circ}\pm$9$^{\circ}$.  The strong variability of its
optical
brightness was first reported by Lu (1972) who monitored it for 
5 years.  Subsequently it was found that its B magnitude had varied 
by $\sim$ 6 magnitudes since 1899.6 (Liller \& Liller 1975). 
{\it The range of brightness spanned is larger
than that known for any other quasar.} Its radio emission exhibits very
rapid, large amplitude variations in both total and polarized flux
(Aller,
Aller \& Hodge 1981; Aller, Aller \& Hughes 1996). It is a core 
dominated radio source with a {\it
one-sided} jet which subtends about 8\farcs0~in the 20-cm waveband 
(O'Dea, Barvainis \& Challis 1988). 
Strong infrared (see Landau \etal 1986), millimeter (see Steppe \etal
1988), 
and UV emission (Malkan \& Moore 1986) have also been detected from 
PKS~1510-089.  $\gamma$-ray emission from PKS~1510-089 has also been 
detected  by the $EGRET$ instrument on {\it CGRO} (Thompson et al 1993, 
Sreekumar \etal 1996) with a flux of (2.3$\pm$0.57)$\times$10$^{-7}$ 
photons cm$^{-2}$ s$^{-1}$ for $\gamma$-rays with energies $\geq$100
MeV, 
and a photon spectral index of 2.51$\pm$0.36.

The broad-band X-ray observations in 1984 and 1985 with 
$EXOSAT$ described by Singh \etal (1990), showed that its X-ray spectrum
is 
best fitted by a power-law with photon index $\Gamma$=1.40$\pm$0.35 and 
low energy absorption consistent with the Galactic value of N$_{\rm H}$ 
in its direction (8.1$\times$10$^{20}$cm$^{-2}$) as given by Stark \etal 
(1992). 
The 2--10~keV flux from $EXOSAT$ observations, was found to be 
8.9$\times$10$^{-12}$ ergs cm$^{-2}$ s$^{-1}$ in the observer's frame. 
The source is highly luminous with its X-ray luminosity alone being 
$\simeq$7$\times$10$^{45}$ ergs s$^{-1}$, assuming isotropic emission,
redshift of 0.361, H$_o$ = 50 km s$^{-1}$ Mpc$^{-1}$, and q$_o$ = 0 in 
the Friedmann cosmology.  

In this letter, we present new $ASCA$ observations of \pks, along 
with an analysis of archival $ROSAT$ data.
Details of the observations and data reduction are given in \S2, 
and the results from the spectral analysis in \S3.  
We discuss our findings in \S4.

\section {OBSERVATIONS AND DATA REDUCTION}
\subsection {$ASCA$}
   PKS~1510-089 was observed with $ASCA$ on 1996 August 20 
as part of the guest observer program (see Table 1 for the log of
observations).
The {\it ASCA} observatory (Tanaka \etal 1994),
contains four imaging thin--foil grazing incidence X-ray telescopes,
two of which are equipped with Solid State Imaging Spectrometers
(SIS) and the other two with Gas Imaging Spectrometers (GIS).  
SIS detectors have been described in detail by Burke et al. (1991),
and GIS by Ohashi et al. (1996) and Makishima et al. (1996).
Their properties have been summarized, e.g., in Singh et al. (1996).
Due to radiation damage since the launch of ASCA, at the
time of the observations reported here the spectral 
resolution of the SIS detectors had been degraded to 
$\sim$10\% at 1~keV.  This spectral resolution is taken into 
account by the response matrix generated as part of the analysis of
this spectrum.

The data were selected as described in Singh et al. (1996) using the
FTOOLS/XSELECT software package. 
\pks~is the only source detected in the field of view.
The counts and pulse height spectra were accumulated from a source
region
of 4 arcmin radius in the SIS and 6 arcmin radius in the GIS, while
the background was taken from the off-axis source-free regions in the
SIS 
and GIS where counts from the outer wings of the point spread function
of 
the source were minimized.  The source spectra were grouped 
to have a minimum of 20 counts per energy channel prior to spectral 
analysis (see \S 3). 
We have compared our results to those obtained assuming 
background spectra accumulated from deep observations of 
blank sky, and obtain consistent results (within statistical
uncertainties).

Steady X-ray emission from PKS~1510-089 was detected with a count rates
of 
$\sim$0.18 and $\sim$0.14 counts s$^{-1}$ above background in SIS0 and
SIS1 detectors and in energy bands of 0.5 -- 8.2 keV and 0.65 -- 8.0 keV
respectively.  Similarly, the count rates observed with GIS2 and
GIS3 were $\sim$0.12 and $\sim$0.14 counts s$^{-1}$ 
in the energy bands of 0.75 -- 9.7 keV and 0.70 -- 10.0 keV
respectively.  
No significant variability was seen on time scales of minutes to 
hours in any of the detectors. 
The average X-ray flux in the 0.4 -- 10 keV energy band in the
observer's 
frame was $\sim$1.1$\times$10$^{-11}$ ergs cm$^{-2}$ s$^{-1}$ which 
implies an X-ray luminosity of $\sim$10$^{46}$ ergs s$^{-1}$ 
(z=0.361; H$_o$ = 50 km s$^{-1}$ Mpc$^{-1}$; q$_o$ = 0 in the Friedmann 
cosmology).  
Similarly, the 2--10~keV flux in the observer's frame, was  
8.6$\times$10$^{-12}$ ergs cm$^{-2}$ s$^{-1}$ which is within 4\% of the
value observed with EXOSAT, consistent
with the uncertainties in the cross-calibration of the 
absolute fluxes of the two satellites.
The flux estimates are based on the best fit spectral models described
in
\S3.

\subsection {$ROSAT$}
PKS~1510-089 was detected in the $ROSAT$ all sky survey (RASS) as well 
as during the pointed observations (Siebert et al. 1995). We have 
retrieved the publicly-available $ROSAT$ data from the pointed 
observations from the archives maintained at NASA High-Energy
Astrophysics
Science Archive Research Center (HEASARC). 
The pointed observations were carried out on 1992 August 17 as part of 
the $ROSAT$ Guest Observer Program, utilizing a position sensitive 
proportional counter (PSPC) as the focal plane detector (Tr\"{u}emper
1983; 
Pfeffermann \etal  1987).  The PSPC has a field of view of 
2\degrees, an energy resolution (FWHM) of 0.42\% at 1 keV and a nominal 
(gain-sensitive) bandwidth for $>$10\% efficiency of 0.1--2.0~keV.
No other X-ray source was detected in the neighborhood of \pks.  
The details of the observations are given in Table 1. 
The on-source counts were selected from a 
circular region having a radius of about 3.25~arcmin and
centered on the source. An X-ray spectrum from the region was
accumulated 
for the entire observation. The background was accumulated from 
several neighboring regions at approximately the same offset as the
source.  
The spectral data from the source were binned from the original
256 channels by grouping the first 176 channels by 8 and the rest by 16
to
improve the signal--to--noise ratio in each energy bin.
These observations resulted in a reasonably good quality spectra
with typical signal--to--noise ratio per bin of 5$\sigma$. 

Steady soft X-ray emission from PKS~1510-089 was detected with a count
rate 
of $\sim$0.18 counts s$^{-1}$ above background in the energy range
of 0.13 -- 2.0 keV in the observer's frame.
According to Siebert \etal (1995), the source flux changed by a factor 
of $\sim$3.5 between the RASS and the pointed mode observation 
about 2 years later.  The average X-ray flux in the 1992 pointed 
mode observation in the 0.1 -- 2 keV energy band was 
2.1$\times$10$^{-12}$ ergs cm$^{-2}$ s$^{-1}$ in the observer's frame.
The flux estimate is based on the best fit spectral models in \S3.

\section {SPECTRAL ANALYSIS AND RESULTS}
\subsection {$ASCA$}
We have analyzed the X-ray spectra obtained from SIS data for E$\geq$0.5
keV 
and the GIS data for energies $\geq$0.7 keV.  
The spectra were analyzed using \verb+XSPEC+ v9.0 (Arnaud 1996).
We first analyzed the data from each detector separately using a simple 
absorbed power-law model with absorption cross-sections and abundances
from Morrison \& McCammon (1983).  
The spectra SIS0, SIS1, GIS2 and GIS3
all gave consistent results.  The analysis was then performed jointly on
the
2 SIS detectors, and on the 2 GIS detectors; and finally on all the four
detectors jointly. We fit the spectra from all four instruments 
simultaneously, but allowing the normalization of the model to be free
for 
each detector to account for differences in the absolute flux
calibrations 
of the instruments.  The results based on the joint analyses of all 4
detectors  are given in Table 2.  The results are consistent with the 
analysis of spectra from individual detectors, analyzed separately.  
The observed spectra from the 4 detectors along with the best fit model
of an 
absorbed power-law are shown separately for SIS detectors in Figure 1,
and 
for GIS detectors in Figure 2. The fit is very good, having a reduced 
chi-square statistic, \chinu~= 1.03. 
The resulting spectrum is hard with a photon
index $\Gamma$=1.30$^{+0.06}_{-0.06}$
(where the errors are at 90\% confidence for one parameter of interest),
and an equivalent hydrogen column density estimated from low-energy 
absorption of $N_{H}=16\pm4\times$\NHunits 
which is about twice that due to absorption in our galaxy (see \S1). 
The same result is obtained if assume an  absorber at z=0.361, along
with a
fixed absorption from 21-cm measurements in our galaxy.
In Figure 3 we show the 68\%, 90\%, and 99\% confidence contours for the
values of the spectral index and the \NH~allowed by the data.  
Motivated by an earlier report about the presence of a Gaussian feature
near
5 keV in the spectrum (Singh et al. 1990), 
we also tried adding a similar feature to the $ASCA$
spectra and got a marginal improvement in the quality of the fit. 
The best fit line position was at 5.1$\pm$0.4 keV with an equivalent 
width of about 87 eV. 
Considering the signal to noise in this portion of the spectrum and the
fact
that the overall fit improved only marginally, we regard this equivalent
width as an upper limit. The equivalent width of the emission line 
suggested by an $EXOSAT$ observation (Singh et al 1990) was several
times
higher, suggesting that either the line varied or it was spurious.
Alternate models e.g., (i) a broken power-law with a low-energy index 
$\Gamma_l$, a break at energy E$_c$ $\sim$0.9 keV, and a high-energy
index
$\Gamma_h$, or (ii) the addition of a blackbody to a simple power-law,
or (iii) a partially covered power-law,  provided
equally good fits to the spectra (see Table 2).  
The improvement in the fit was insignificant, however.  Therefore,
the data are not able to distinguish between the absence or the presence of 
a low-energy excess in the spectrum.

\subsection {$ROSAT$}
Results from our spectral analysis of the {\it ROSAT} PSPC data are 
presented in Table 3.
A simple absorbed power-law model provided a good fit to the PSPC
spectrum.  
The data and our best fit power-law model are shown in Figure 4.  
The confidence contours for the spectral index and the hydrogen
column density \NH~allowed by the PSPC data are shown in Fig. 3.
The best fit spectral index, $\Gamma$=1.88$\pm$0.28 is
significantly higher than that measured during the $ASCA$ 
observations.  The corresponding column density \NH, is 
(7.6$^{+2.0}_{-1.5}$)$\times$\NHunits, which is consistent with 
the 21-cm value.  The unabsorbed flux at 1 keV is $\sim$10\% lower 
than in the $ASCA$ observations.
The apparent discrepancy between the $ROSAT$ and $ASCA$
observations imply that either variations in both the
spectral index and the absorbing column occurred between the
two epochs, or that the PSPC band is dominated by a 
separate, steep spectral component.
The PSPC spectra could be fit equally well by a broken power-law
($\Gamma_l$=
1.9$\pm$0.2, with the hard component fixed at $\Gamma_h$=1.30) with
absorption, however the break energy cannot be constrained (see Table
3.). 
A model consisting of a low-energy blackbody and 
hard power-law also provides a statistically acceptable 
description of the data. However, such a model requires 
a value of \NH~less than the Galactic value and hence we consider
it unrealistic.  The PSPC data show no preference for any of the three
models.

\subsection{$ROSAT$ and $ASCA$ Combined Spectral Analysis}
If we accept the simple absorbed power-law then there
is a significant difference between the spectral index and \NH~observed
by $ROSAT$ in 1992 compared to that observed by $ASCA$ 4 years later. 
Since there were no measurements above 2 keV in 1992, the hard X-ray 
flux seen with $ASCA$ might have varied even though the 1-keV flux 
hardly changed.  In that case, joint fitting of $ROSAT$ and $ASCA$ 
data would not be meaningful.  In the likely case that a more complex
spectrum is present with a change around 1-keV, the joint
fitting can constrain the additional parameters involved in
multi-component
models.  We, therefore, carried out multi-component, multi-instrument
model
fitting, the results of which are given in Table 4.  
We fit the spectra from all five instruments simultaneously, 
but allowing the normalization of the model to be free for each 
detector to account for differences in the absolute flux calibrations
of the instruments and changes in source intensity between the two
epochs.  
As expected, a simple power-law model gives
an acceptable fit only if the \NH~is allowed to be significantly lower
than the Galactic value in the direction of \pks.   
Both the broken power-law and blackbody plus power-law models 
give acceptable fits to the data with the \NH~value almost 
consistent with the Galactic value (see Table 4). 
A partially covered power-law model with Galactic absorption and a very 
high absorption intrinsic to the source at z=0.361, gives a somewhat
poorer fit.  Thus we conclude that the most likely explanation of the 
$ROSAT$ \& $ASCA$ observations is that a powerlaw of photon index 
$\Gamma \simeq$ 1.3 dominates the spectrum above $\gtrsim$ 1~keV with 
a 2--10~keV flux $\simeq$8.6$\times$10$^{-12}$ ergs cm$^{-2}$ s$^{-1}$ 
at both epochs in the observer's frame. 
A separate, steeper spectral component dominates at lower energies, 
the specific form of which cannot be unambiguously determined.

\section {DISCUSSION}
$ASCA$ observations show that the 0.5 -- 10 keV X-ray spectrum of 
\pks~is hard ($\Gamma \simeq$ 1.3).   Although earlier observations with 
the $EXOSAT$ (Singh \etal 1990; Sambruna \etal 1994) had indicated
that the X-ray spectrum of \pks~might be this hard,
those measurements were not conclusive due to a relatively 
poor signal-to-noise ratio.
The observed spectral index is significantly flatter than that
of the typical radio-loud quasars ($< \Gamma >$=1.66$\pm$0.07,
Lawson \etal 1992; Cappi \etal 1997), 
radio-quiet quasars($< \Gamma >$=1.90 $\pm$0.11, Lawson \etal 1992; 
Williams \etal 1992), 
and BL Lac objects ($< \Gamma >$ = 2.20$^{+0.17}_{-0.15}$,
Sambruna \etal 1994).
The observed spectrum is, however, similar to that of other HPQs 
observed in both the hard 
(e.g. 3C~345 with $\Gamma$=1.4$\pm$0.09, Makino 1989; 
PKS~0537-441 with $\Gamma$=1.26$^{+0.24}_{-0.31}$ Sambruna et al. 1994)
and the soft
(e.g. 0212+735 with $\Gamma$=0.44$^{+0.39}_{-0.41}$ 
and 0836+710 with $\Gamma$=1.4$\pm$0.05, Brunner \etal 1994)
X-ray bands.
Thus, based on an admittedly small number of examples, 
HPQs do appear to show X-ray spectra that are flatter than those of
other core-dominated radio-loud quasars.   

PKS 1510-089 was observed in the near ultraviolet band with the
International
Ultraviolet Explorer (IUE) low-dispersion spectrographs during the early
1980s. There are three SWP exposures and two LWR exposures in the
archive. 
The SWPs include one nondetection and two spectra obtained over a
nominal 
1-year baseline whose continuum levels are equivalent to the 
1$\sigma$ level. One LWR exposure was underexposed and resulted in 
a nondetection. There is, however, no evidence of variability in the 
available UV data, although it cannot be ruled out either. 
Though weakly detected on one occasion with the short- and long-
wavelength instruments, redshifted emission lines of Ly-$\alpha$ and 
O~VI/Lyman-$\beta$ are clearly detected, with several other normally
prominent
emission lines possibly present. The UV continuum is well approximated
by a
power law with index -1.2 ($\nu$ vs $f_{\nu}$). 

We constructed a multifrequency spectrum using measurements
from the published literature.  The spectrum is shown in Figure 5.
We note that these data do not result from simultaneous or even 
contemporaneous observations. The ``radio'' points at 1.5 and 4.9 GHz
are 
from O'Dea et al. (1985); we note however that Aller et al. (1996), who 
present a 20-year radio light curve for PKS 1510-089, show that 
it varies by at least factors of 5, mainly in series of 
flare-like events occurring on ~yearly time scales (and sometimes 
persisting for ~months). The near IR measurement at 1.5$\mu$m 
is from Landau et al. (1986) and we used the visual magnitude of 
V=16.5 of Bolton \& Elkers (1966) to derive  a
flux point corresponding to 5500 \AA. We used the UV measurements 
obtained from the IUE archives as described previously.  
The optical and UV fluxes have been corrected for galactic extinction 
using the extinction law of Seaton (1979) and a color excess estimated 
from the galactic \NH~values derived from the 21-cm measurements.
The $\gamma$-ray flux at energies $\geq$100 MeV has been
taken from $EGRET$ measurements (Thompson et al 1993).

The multiwaveband spectrum of 
PKS~1510-089 resembles that of many other blazars detected by $EGRET$, 
with the bulk of the luminosity being emitted as $\gamma$-rays
(e.g. see Fig.5 of von Montigny et al (1995) -- indeed, we note the 
multiwaveband spectrum of PKS~1510-089 is globally very similar to the 
BL Lac Mrk~421).
In the popular relativistic jet models, the hard X- and $\gamma$-ray 
emission is proposed to arise as a result of inverse-Compton scattering
of 
low energy photons by relativistic electrons within the jet
(e.g., see von Montigny et al (1995), and references therein).
We suggest that the double powerlaw form found above might indicate 
that the {\it ROSAT} data are dominated by the steep tail of this low
energy
emission (often assumed to be synchrotron radiation) while the 
flatter inverse-Compton scattered component dominates
the {\it ASCA} data $>1$~keV.
We suggest that even though superluminal motion has not (yet) been
detected 
in this source, the emission is likely to be relativistically beamed.

     Based on Figure 5, PKS 1510-089 may be among a subclass 
of blazar AGN which exhibit the so called big blue-bump component.
In order to study the blue-bump emission, we subtracted a 
power-law  component, connecting the X-ray, IR, and mm radio 
points, from the multifrequency spectrum to obtain residuals in 
the optical-UV region. A statistically significant, positive 
residual is obtained. We emphasize that variability could 
account for some or all of this apparent residual.  We then fit a
thermal
accretion disk spectrum to these residual spectra.  The model 
involves thermal emission due to viscous dissipation in a 
steady-state, accretion disk which has been modified to include 
the effects of gravitational red-shift and focusing (e.g. Sun 
\& Malkan, 1989); we have employed computational methods similar
to those described in Czerny et al. (1986).  A Schwarzchild 
metric was used in the calculations and an inclination of 
sin~$i$ = 0.5 was assumed.  We fixed the inner disk radius at 3R$_{g}$ 
and R$_{out}$/R$_{in}$$\simeq$200. We were unable to obtain a reasonable
fit
to the optical and UV data without reducing R$_{out}$/R$_{in}$ to this 
level; values more typically fitted to AGN are of order  
R$_{out}$/R$_{in}$$\simeq$10$^3$.  Again, this could possibly be
attributed to variability -- we suspect that the actual visual flux at 
the time of the UV measurements may have been significantly higher. 

    If the blue-bump residual we depict here is NOT an artifact 
of non-simultaneous measurements of a highly variable source, 
PKS 1510-089 is among a minority of the extreme blazar AGN which 
both emit gamma-rays and exhibit a blue-bump component. 
Another blazar with a blue-bump 
component is 3C 345 (e.g. Webb et al 1994), but it was
never detected at more than a 3.5$\sigma$ level with $EGRET$. 

A detailed discussion of the various physical models proposed for such 
objects is beyond the scope of the current paper. 
Further simultaneous, broad-band UV, X- \& $\gamma$-ray observations are
required to constrain physical models.

\acknowledgments
We wish to thank the entire $ASCA$ team for making these observations
possible.
This research has made use of $ROSAT$ archival data obtained through the
High Energy Astrophysics Science Archive Research Center, HEASARC,
Online
Service, provided by the NASA-Goddard Space Flight Center.

\clearpage
\begin{figure}[h]
 
\figurenum{1}
\caption
{Best-fit SIS spectra from a joint fit to the SIS and GIS spectra are shown
after fitting with a single absorbed power-law (upper panel).  
The best fit model is shown as a histogram.  The departures from the best fit
are shown in the lower-panel as the number of $\sigma$s with an error bar of
1 $\sigma$.}

\figurenum{2}
\caption
{Same as in Figure 1, but for the GIS spectra.}

\figurenum{3}
\caption
{Allowed ranges of the N$_H$ and $\Gamma$ from the X-ray spectrum 
of PKS~1510-089 based on the \chisq~contours. The full contours represent 
the 68\%, 90\%, and 99\% confidence limits from fitting SIS+GIS together, 
the dashed-dotted contours show similar confidence levels for the SIS data
alone, the dotted contours for the GIS data alone, and the dashed contours  
are for the PSPC data alone. The plus, cross, cross, and plus mark the bext fit
values.}

\figurenum{4}
\caption
{Best-fit PSPC spectra using a single absorbed power-law, shown in the
same way as for the Figs. 1 and 2.}  

\figurenum{5}
\caption
{Broadband  spectral energy distribution of PKS 1510--089 based on the 
(non-contemporaneous) multifrequency measurements cited in the text. 
The dotted line represents a powerlaw fit to the mm, IR and X-ray 
measurements, with a superposed multi-color blackbody disk fit to the 
apparent blue-bump residual. The outer disk radius required to obtain 
the fit was much smaller than is normally the case for radio quiet 
AGN -- we suspect that the actual optical flux at the time 
of the UV measurements was probably higher.} 

\end{figure}
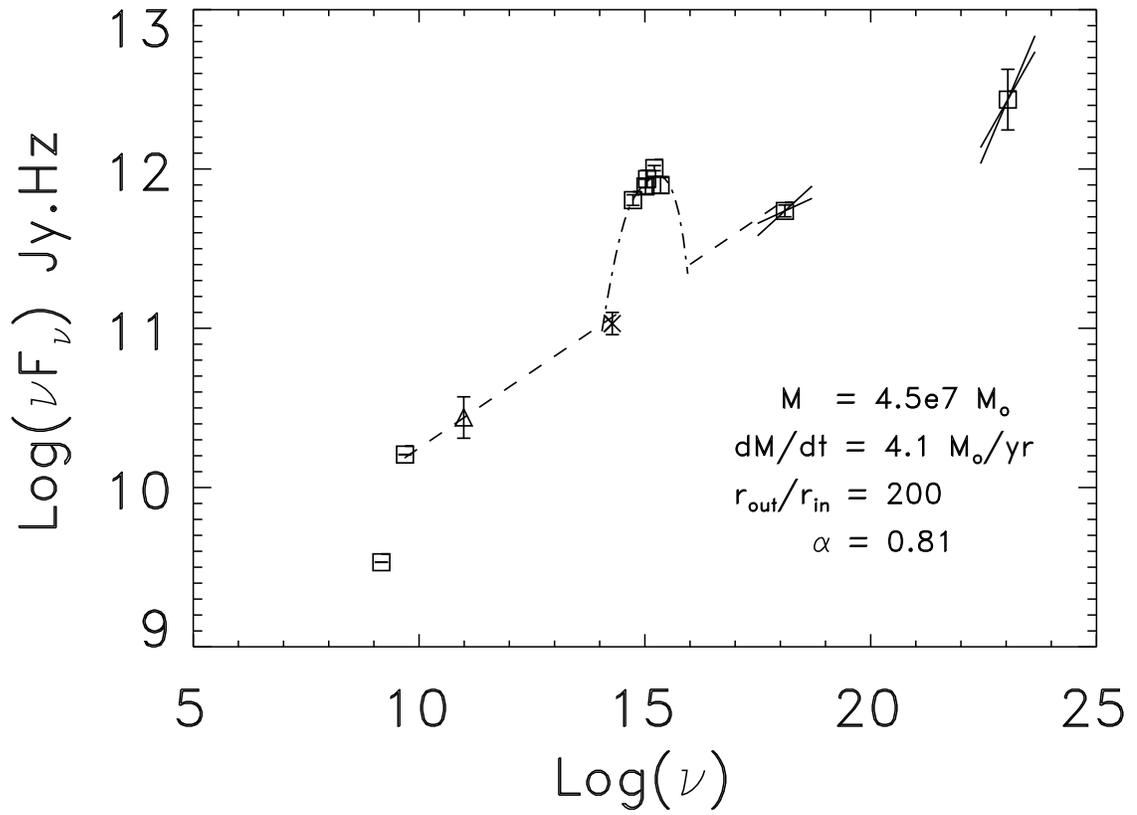
\end{document}


\begin{planotable}{ccrrrc}
\tablenum{1}
\tablecaption{OBSERVATION LOG FOR PKS~1510-089} 
\tablewidth{40pc}
\tablehead{
\colhead{Satellite} &
\colhead{Detector} &
\colhead{Date } &
\colhead{Exposure} &
\colhead{Count Rate\tablenotemark{a}} &
\colhead{Energy Range\tablenotemark{b}}\\  
\colhead{Name} &
\colhead{Name} &
\colhead{} &
\colhead{(seconds)} &
\colhead{(Counts s$^{-1}$)} &
\colhead {(keV)} 
}
\startdata
ROSAT     &  PSPC  & 1992 Aug 17 & 5153  & 0.176$\pm$0.006  &  0.13 -- 1.9 \nl 
ASCA      &  SIS0  & 1996 Aug 20 & 18317 & 0.180$\pm$0.004  &  0.51 -- 8.2 \nl
ASCA      &  SIS1  & 1996 Aug 20 & 18105 & 0.141$\pm$0.003  &  0.65 -- 8.0 \nl
ASCA      &  GIS2  & 1996 Aug 20 & 18301 & 0.118$\pm$0.003  &  0.75 -- 9.7 \nl
ASCA      &  GIS3  & 1996 Aug 20 & 18300 & 0.138$\pm$0.003  &  0.70 -- 10.0 \nl
\enddata
\tablenotetext{a}{After background subtraction.}
\tablenotetext{b}{In observer frame.}
\end{planotable}


\begin{planotable}{lcccccccc}
\tablenum{2}
\tablewidth{45pc}
\tablecaption{Results from analysis of SIS \& GIS Spectra}
\tablehead{
\colhead{Model} & 
\colhead{\NH} & 
\colhead{$\Gamma_h$} &  
\colhead{A\tablenotemark{a}} & 
\colhead{$\Gamma_l$} & 
\colhead{E$_{bk}$\tablenotemark{d}} & 
\colhead{\chisq/dof} & 
\colhead{E$_l$} & 
\colhead{EW} \\
\colhead{} & 
\colhead{} & 
\colhead{} & 
\colhead{} &
\colhead{or} &  
\colhead{or} &
\colhead{} & 
\colhead{} & 
\colhead{} \\
\colhead{} & 
\colhead{\NHunits} & 
\colhead{} & 
\colhead{} & 
\colhead{kT$_{bb}$\tablenotemark{b}} & 
\colhead{L$_{bb}$\tablenotemark{d}} & 
\colhead{} & 
\colhead{keV} & 
\colhead{eV} \\  
\colhead{} & 
\colhead{} & 
\colhead{} & 
\colhead{} & 
\colhead{or} & 
\colhead{} & 
\colhead{} & 
\colhead{} & 
\colhead{} \\  
\colhead{} & 
\colhead{} & 
\colhead{} & 
\colhead{} & 
\colhead{pcf\tablenotemark{c}} & 
\colhead{} & 
\colhead{} & 
\colhead{} & 
\colhead{}   
}
\startdata
 Power-law & 16.0$^{+4.0}_{-3.7}$ & 1.30$^{+0.06}_{-0.06}$ & 11.2 & -- & -- & 458.3/444 &  -- & -- \nl  
 Power-law & 16.6$^{+3.4}_{-3.6}$ & 1.32$^{+0.03}_{-0.05}$ & 11.4 & -- & -- & 453.5/442 &  5.1$\pm$0.5 & 87 \nl
 +Gaussian  & & & & & & & & \nl
 Broken    & 17.2$^{+6.1}_{-4.8}$ & 1.31$^{+0.07}_{-0.06}$ & 10.8 & 1.8$^{+1.1}_{-1.4}$ & 0.9 & 458.1/442 &  -- & -- \nl
 Power-law  & & & & & & & &  \nl
 Broken & 18.0$^{+5.0}_{-3.0}$ &  1.33$^{+0.06}_{-0.05}$ & 11.7 & 1.8$^{+1.3}_{-1.3}$ & 0.9 & 453.1/440 & 5.1$\pm$0.5 & 88 \nl
 Power-law  & & & & & & & &  \nl
 +Gaussian  & & & & & & & &  \nl
 Power-law & 11.3$^{+5.0}_{-3.5}$ & 1.07$^{+0.17}_{-0.10}$ & 7.6 & 0.45 & $<$1.0 & 452.2/442 &  -- & -- \nl
 +Blackbody & & & & & & & &  \nl 
 Partially & 32$^{+80}_{-25}$ & 1.31$^{+0.08}_{-0.07}$ & 11.3 & 0.54$^{+0.46}_{-0.31}$ & -- & 458.4/443 & -- & -- \nl
 covered       & & & & & & & & \nl 
 Power-law$^{\rm e}$ & & & & & & & & \nl 
\enddata
\tablenotetext{a}{Unabsorbed flux at 1 keV (observer frame) in units of 10$^{-4}$ photons cm$^{-2}$ s$^{-1}$ keV$^{-1}$.}
\tablenotetext{b}{$\Gamma_2$: Low-energy photon index; kT$_{bb}$: Blackbody temperature in keV.}
\tablenotetext{c}{pcf: partially covering fraction of the source.}
\tablenotetext{d}{E$_{bk}$: Break energy in keV for the broken power-law; L$_{bb}$: Total Blackbody luminosity in 10$^{44}$ergs s$^{-1}$.}
\tablenotetext{e}{Fixed \NH = 8.1 $\times$ \NHunits due to galaxy + absorption at z =0.361.}
\tablenotetext{} {Notes: Errors and upper limits are with 90\% confidence based on \chisq$_{\rm min }$+2.71}

\end{planotable}

\newpage

\begin{planotable}{crcccccccc}
\tablenum{3}
\tablewidth{40pc}
\tablecaption{Results from Analysis of PSPC Spectrum}
\tablehead{
\colhead{S.} &
\colhead{Model} & 
\colhead{\NH} & 
\colhead{$\Gamma_h$} & 
\colhead{ A\tablenotemark{a}} & 
\colhead{$\Gamma_l$} & 
\colhead{E$_{bk}$\tablenotemark{c}} & 
\colhead{\chisq/dof} & 
\colhead{F$_{SX}$\tablenotemark{d}} &  
\colhead{L$_{SX}$\tablenotemark{e}} \\
\colhead{} & 
\colhead{} & 
\colhead{} & 
\colhead{} & 
\colhead{} &
\colhead{or} &  
\colhead{or} &
\colhead{} & 
\colhead{} \\
\colhead{No.} &
\colhead{} &
\colhead{\NHunits} & 
\colhead{}& 
\colhead{}& 
\colhead{kT$_{bb}$\tablenotemark{b}} & 
\colhead{L$_{bb}$\tablenotemark{c}} & 
\colhead{} & 
\colhead{} & 
\colhead{}  
}
\startdata
1. & Power-law & 7.6$^{+2.0}_{-1.5}$ & 1.88$\pm$0.28 & 10.2 & -- & -- & 25.9/20 & 2.1 & 3.5 \nl 
2. & Broken & 7.7$^{+2.0}_{-1.7}$ & 1.30 & 10.2 & 1.9$\pm$0.2 & 1.6$^{+\inf}_{-0.4}$ & 25.6/19 & 2.1 & 3.6 \nl 
   & Power-law & & & & & & & & \nl
3. & Power-law & 5.1$^{+1.5}_{-1.3}$ & 1.30 & 6.2 & 0.21$\pm$0.05 & 7.0$\pm$4.0 & 24/18 & 2.0 & 2.2 \nl 
   & +Blackbody & & & & & & & & \nl
\enddata
\tablenotetext{a}{Unabsorbed flux at 1 keV (observer frame) in units of 10$^{-4}$ photons cm$^{-2}$ s$^{-1}$ keV$^{-1}$.}
\tablenotetext{b}{$\Gamma_2$: Low-energy photon index; kT$_{bb}$: Blackbody temperature in keV.}
\tablenotetext{c}{E$_{bk}$: Break energy in keV for the broken power-law; L$_{bb}$: Total Blackbody luminosity in 10$^{44}$ergs s$^{-1}$.}
\tablenotetext{d}{Absorbed flux between 0.1 -- 2.0 keV (observer frame) in units of 10$^{-12}$ ergs cm$^{-2}$ s$^{-1}$.}
\tablenotetext{e}{Intrinsic Luminosity (no absorption) between 0.1 -- 2.0 keV (observer frame) in units of 10$^{45}$ergs s$^{-1}$.}
\tablenotetext{} {Notes: Errors and upper limits are with 90\% confidence based on \chisq$_{\rm min }$+2.71}

\end{planotable}


\begin{planotable}{lcccccc}
\tablenum{4}
\tablewidth{40pc}
\tablecaption{Results from joint analysis of PSPC, SIS \& GIS Spectra}
\tablehead{
\colhead{Model} & 
\colhead{\NH} & 
\colhead{$\Gamma_h$} &  
\colhead{A\tablenotemark{a}} & 
\colhead{$\Gamma_l$} & 
\colhead{E$_{bk}$\tablenotemark{d}} & 
\colhead{\chisq/dof} \\ 
\colhead{} & 
\colhead{} & 
\colhead{} & 
\colhead{} &
\colhead{or} &  
\colhead{or} & 
\colhead{} \\
\colhead{} & 
\colhead{\NHunits} & 
\colhead{} & 
\colhead{} & 
\colhead{kT$_{bb}$\tablenotemark{b}} & 
\colhead{L$_{bb}$\tablenotemark{d}} & 
\colhead{} \\ 
\colhead{} & 
\colhead{} & 
\colhead{} & 
\colhead{} & 
\colhead{or} & 
\colhead{} & 
\colhead{} \\ 
\colhead{} & 
\colhead{} & 
\colhead{} & 
\colhead{} & 
\colhead{pcf\tablenotemark{c}} & 
\colhead{} & 
\colhead{} 
}
\startdata
 Power-law & 5.1$^{+1.1}_{-0.7}$ & 1.17$^{+0.04}_{-0.03}$ & 9.3 & -- & -- & 531.2/466   \nl  
 Broken    & 11.3$^{+3.0}_{-3.0}$ & 1.24$^{+0.04}_{-0.04}$ & 4.9 & 3.5$^{+0.0}_{-1.2}$ & 0.70$^{+0.10}_{-0.10}$ & 520.7/464    \nl
 Power-law &  & & & & & \nl
 Power-law & 11.8$^{+2.0}_{-1.6}$ & 1.25$^{+0.05}_{-0.05}$ & 10.4 & 0.056$^{+0.01}_{-0.01}$ & 7.6 & 520.3/464  \nl
 +Blackbody & & & & & &  \nl 
 Partially & 200$^{+220}_{-50}$ & 1.30$^{+0.17}_{-0.05}$ & 11.3 & 0.15$^{+0.75}_{-0.12}$ & --  & 538.1/465  \nl
 covered & & & & & &   \nl 
 Power-law$^{\rm e}$ & & & & & &  \nl 
\enddata
\tablenotetext{a}{Unabsorbed flux at 1 keV (observer frame) in units of 10$^{-4}$ photons cm$^{-2}$ s$^{-1}$ keV$^{-1}$.}
\tablenotetext{b}{$\Gamma_2$: Low-energy photon index; kT$_{bb}$: Blackbody temperature in keV.}
\tablenotetext{c}{pcf: partially covering fraction of the source.}
\tablenotetext{d}{E$_{bk}$: Break energy in keV for the broken power-law; L$_{bb}$: Total Blackbody luminosity in 10$^{45}$ergs s$^{-1}$.}
\tablenotetext{e}{Fixed \NH = 8.1 $\times$ \NHunits due to galaxy + absorption at z =0.361.}
\tablenotetext{} {Notes: Errors and upper limits are with 90\% confidence based on \chisq$_{\rm min }$+2.71}

\end{planotable}